\documentclass[preprint,superscriptaddress,preprintnumbers,amsmath,amssymb,prb]{revtex4}

\usepackage{graphicx}
\usepackage{color}
\usepackage[amssymb]{SIunits}

\begin{document}

\title{Comparison of gold- and graphene-based resonant nano-structures for terahertz metamaterials and an ultra-thin graphene-based modulator}

\author{Nian-Hai Shen}
\email[]{nhshen@ameslab.gov}
\affiliation{Ames Laboratory---U.S.~DOE and Department of Physics and Astronomy, Iowa State University, Ames, Iowa 50011, USA}

\author{Philippe Tassin}
\affiliation{Ames Laboratory---U.S.~DOE and Department of Physics and Astronomy, Iowa State University, Ames, Iowa 50011, USA}
\affiliation{Department of Applied Physics, Chalmers University, SE-412 96 G$\ddot{o}$teborg, Sweden}

\author{Thomas Koschny}
\affiliation{Ames Laboratory---U.S.~DOE and Department of Physics and Astronomy, Iowa State University, Ames, Iowa 50011, USA}

\author{Costas M. Soukoulis}
\email[]{soukoulis@ameslab.gov}
\affiliation{Ames Laboratory---U.S.~DOE and Department of Physics and Astronomy, Iowa State University, Ames, Iowa 50011, USA}
\affiliation{Institute of Electronic Structure and Laser, FORTH, 71110 Heraklion, Crete, Greece}

\begin{abstract}

Graphene exhibits unique material properties and in electromagnetic wave technology, it raises the prospect of devices miniaturized down to the atomic length scale. Here we study split-ring resonator metamaterials made from graphene and we compare them to gold-based metamaterials. We find that graphene's huge reactive response derived from its large kinetic inductance allows for deeply subwavelength resonances, although its resonance strength is reduced due to higher dissipative loss damping and smaller dipole coupling. Nevertheless, tightly stacked graphene rings may provide for negative permeability and the electric dipole resonance of graphene meta-atoms turns out to be surprisingly strong. Based on these findings, we present a terahertz modulator based on a metamaterial with a multi-layer stack of alternating patterned graphene sheets separated by dielectric spacers. Neighbouring graphene flakes are biased against each other, resulting in modulation depths of over $75\%$ at a transmission level of around $90\%$.

\end{abstract}

\pacs{81.05.Xj, 78.67.Wj, 41.20.Jb}

\maketitle

\section{Introduction}

Since the emergence of graphene \cite{Novoselov2004Science,Novoselov2005Nature} through exfoliation of graphite, it has been shown to exhibit many unique mechanical, thermal, electric and magnetic properties, turning graphene into a prosperous research field. \cite{Geim2007NatMater, Bonaccorso2010NatPhoton, Novoselov2012Nature, Grigorenko2012NatPhoton} Metamaterials---artificial  materials designed towards wave manipulation at the subwavelength scale \cite{Pendry1999IEEE}---have benefited both academia and industry, providing many interesting and valuable  possibilities, \cite{Liu2011CSR,Soukoulis2007Science,Engheta2007Science,Soukoulis2011NatPhoton,Zheludev2010Science,Boltasseva2011Science} such as super-resolution imaging, \cite{Pendry2000PRL,Fang2005Science} cloaking, \cite{Schurig2006Science} energy harvesting , \cite{Landy2008PRL,Hess2007Nature}sensing,  \cite{Liu2009NanoLett}and terahertz (THz) wave manipulation. \cite{Chen2006Nature} Some efforts have been made to take advantage of graphene in the design of metamaterial structures and devices, leading to some initial promising achievements. \cite{Vakil2011Science,Ju2011NatNano,Yan2012NatNano,Lee2012NatMater,Thong2012PRL,Chen2012Nature,Fei2012Nature} Comparing to the optical frequency band, the THz domain may provide an attractive platform for graphene to achieve desirable applications in the scope of metamaterials.  \cite{Tassin2012NatPhoton, Tassin2013Science} In this article, we compare the performance of metamaterials made out of patterned sheets of graphene versus gold.  In this way, we can investigate whether graphene has superior properties over gold to create deep sub-wavelength and strong electromagnetic resonances. In addition, we present a THz device in which the tunable electrical properties of graphene provide unprecedented tunability of a metamaterial resonance, which is very interesting for THz modulation.

\section{Data of graphene and gold}

In view of the importance of using accurate experimental data to describe the electric response of graphene,\cite{Tassin2013Science} we briefly review the data we have used in this study to assess the performance of graphene-based materials and devices. In the terahertz band, the linear response of graphene can be well described by a Drude model through the following dynamic sheet conductivity:\cite{Ju2011NatNano,Yan2011ACSNano}
\begin{equation}
\sigma_\mathrm{s} = \frac{\alpha}{\gamma - \mathrm{i}\omega},
\end{equation}
where $\alpha$ is the Drude weight with unit of \unit{}{\reciprocal\ohm\reciprocal\second}, $\gamma$ represents the collision frequency, which is related to scattering time $\tau$ by $\gamma=1/\tau$, and $\omega=2 \pi f$ is the angular frequency. A more natural and intuitive way to understand and predict the electromagnetic properties of a conductor is by considering the surface impedance $Z_\mathrm{s} = 1/\sigma_\mathrm{s}$:\cite{Tassin2012NatPhoton}
\begin{equation}
Z_\mathrm{s} = \frac{\gamma}{\alpha} - \mathrm{i}\frac{\omega}{\alpha} = R_\mathrm{s} - \mathrm{i}X_\mathrm{s},
\end{equation}
in which the real part (the sheet resistance) is a measure of dissipative loss, whereas the imaginary part (the sheet reactance) characterizes the \textit{kinetic inductance} $L_\mathrm{k}=1/\alpha$.

A widely adopted theoretical data set for the dynamic conductivity of graphene\cite{Papa2013Light} has $(\alpha, \gamma)= (\unit{5.93\times10^{10}}{\reciprocal\ohm\reciprocal\second}, \unit{1.98\times10^{12}}{\reciprocal\second})$ with an effective scattering time $\tau$ of \unit{0.5}{\pico\second} (we will further on refer to this data set as Papasimakis~\textit{et~al.} graphene). The dissipative loss (resistance) for this data set is \unit{33.4}{\ohm}. In the past few years, great efforts have been undertaken towards improving the quality of graphene with fairly low loss and several direct experimental measurements of the terahertz conductivities have become available. Yan~\textit{et~al.} have fabricated high-quality, highly-doped graphene by chemical vapor deposition followed by a chemical doping process to increase the doping level.\cite{Yan2012NatNano} These graphene samples are described by a Drude model [Eqs.~(1) and (2)] with parameters $(\alpha, \gamma)=(\unit{7.6\times10^{10}}{\reciprocal\ohm\reciprocal\second}, \unit{9.8\times10^{12}}{\reciprocal\second})$ with $\tau$ approximately \unit{0.1}{\pico\second} (further on denoted by Yan~\textit{et~al.} graphene). The corresponding dissipative loss is \unit{129}{\ohm}, more than most of theoretical models predict, but still a great improvement compared to previously reported experimental data. We also extracted the conductivity data of graphene from one of the first measurements of the infrared conductivity by Li and co-workers \cite{Li2008NatPhys} and obtained the data set as $(\alpha, \gamma)= (\unit{1.99\times10^{10}}{\reciprocal\ohm\reciprocal\second}, \unit{29.4\times10^{12}}{\reciprocal\second})$, below referred to as Li~\textit{et~al.} graphene with dissipative loss of \unit{1477}{\ohm} \cite{Li-graphene-extraction}. In our analysis of THz graphene metamaterials, we will apply these three data sets of graphene to compare their performance to gold-based metamaterials.

\begin{figure}[htb]
 \includegraphics[width=14 cm]{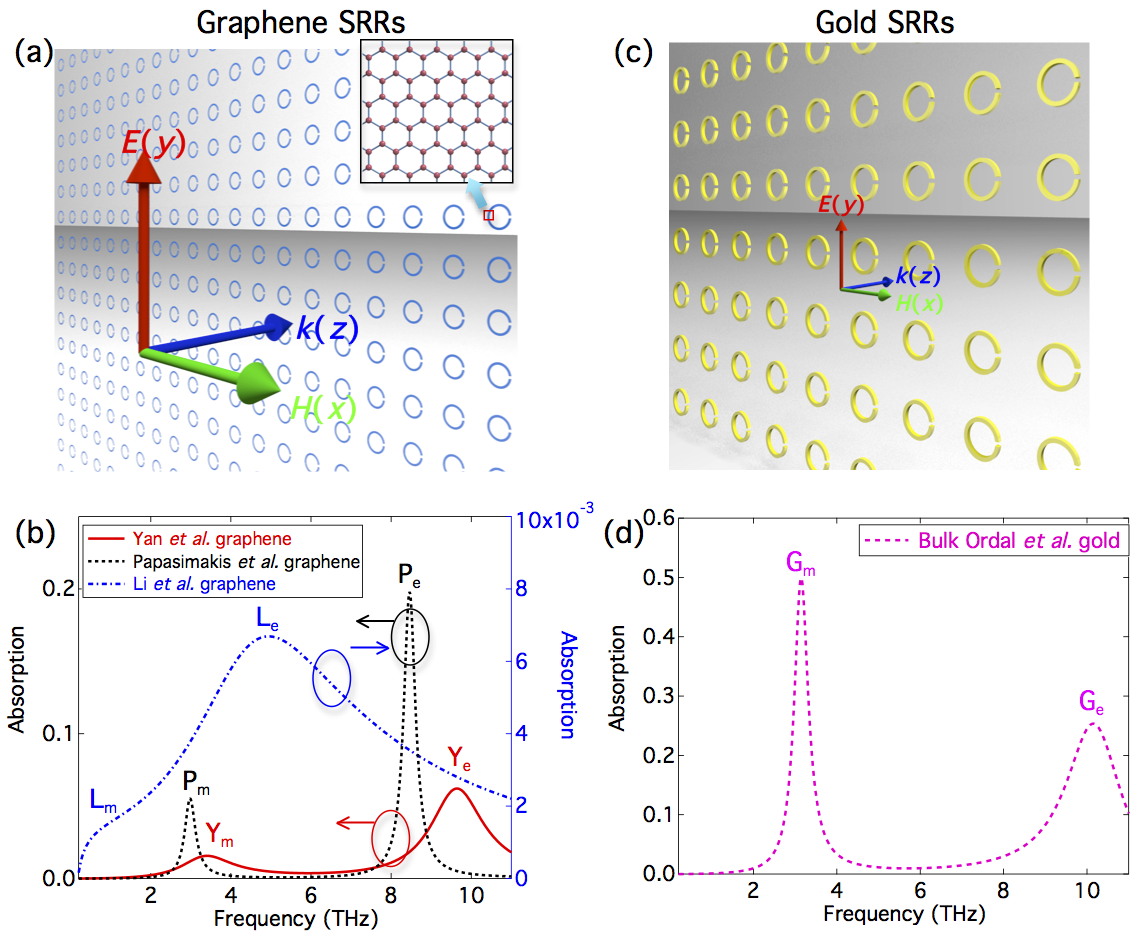}
 \caption{Graphene SRRs [(a)-(b)] and gold SRRs [(c)-(d)] under normal-incidence illumination. 
 (a) Schematic diagram of graphene SRRs. Geometrical parameters are: outer diameter of the ring $D=\unit{1}{\micro\meter}$, ring width $w=100$ nm, gap size $g=100$ nm, and lattice constant of the graphene SRR array $a=\unit{2}{\micro\meter}$. (b) Absorption spectra for SRRs made from Yan \textit{et al.} \cite{Yan2012NatNano}, Papasimakis \textit{et al.} \cite{Papa2013Light}, and Li \textit{et al.} \cite{Li2008NatPhys} graphene. (c) Schematic of 30-nm-thick gold SRRs. Geometrical parameters are: outer diameter of the ring $D=\unit{15}{\micro\meter}$, ring width $w=1.25\;\mu$m, gap size $g=1\;\mu$m, and lattice constant of the gold SRR array $a=\unit{25}{\micro\meter}$. (d) Absorption spectrum for Ordal \textit{et al.} \cite{Ordal1983AO} gold SRRs. The inset in (a) illustrates the honeycomb lattice of graphene.}
\end{figure}

For the gold-based metamaterials, we adopt the commonly used experimental data from Ordal~\textit{et~al.},\cite{Ordal1983AO} which are well described at terahertz frequencies by the following Drude model for the bulk conductivity:
\begin{equation}
\sigma = \frac{\epsilon_0 \omega_\mathrm{p}^2}{\gamma-\mathrm{i}\omega},
\end{equation} 
where $\omega = 2 \pi f$ is the angular frequency, $\omega_\mathrm{p} = 2\pi f_\mathrm{p}$ represents the plasma frequency with $f_\mathrm{p} = \unit{2184}{\tera\hertz}$, and $\gamma= \unit{40.5\times10^{12}}{\reciprocal\second}$, corresponding to an effective scattering time $\tau \approx \unit{24.7}{\femto\second}$. Even though we will use the bulk conductivity of gold in our simulations, it is worthwhile to calculate the equivalent complex sheet conductivity of a gold film, since it provides a straightforward and intuitive estimate of the properties of gold compared to the above mentioned graphene data. For a $d=$\unit{30}{\nano\meter}-thick film of gold, we get the equivalent sheet conductivity parameters $\alpha = \epsilon_0 \omega_\mathrm{p}^2 d \approx \unit{5 \times 10^{13}}{\reciprocal\ohm\reciprocal\second}$ and $\gamma = \unit{40.5}{\tera\hertz}$. We can easily obtain the corresponding dissipative sheet resistance of only \unit{0.8}{\ohm}, which is much smaller than that in graphene. We also take note of earlier work that considered metamaterials made out of a patterned one-atom-thick gold film \cite{Papa2013Light}, and some related discussions can be found in the Supplementary Note I.\cite{SM}

Having discussed the material response of graphene and gold, we now start our detailed comparison of graphene-and gold-based metamaterials in the THz range. The split-ring resonator (SRR), a prototype metamaterial element with strong magnetic response, has been intensively studied and played an important role in the metamaterials field because of its potential negative permeability. Here we consider SRR metamaterials under two different directions of illumination, i.e., normal- and parallel-incidence with respect to the rings. For the following numerical studies, we adopt the commercial electromagnetic software package, i.e., CST Microwave Studio, with which, the single-unit-cell-based simulations are performed by applying the periodic boundary conditions settings. The field monitors are set to obtain the electric, magnetic and current distributions at feature frequencies when necessary.

\begin{figure}[htb]
  \includegraphics[width=\textwidth]{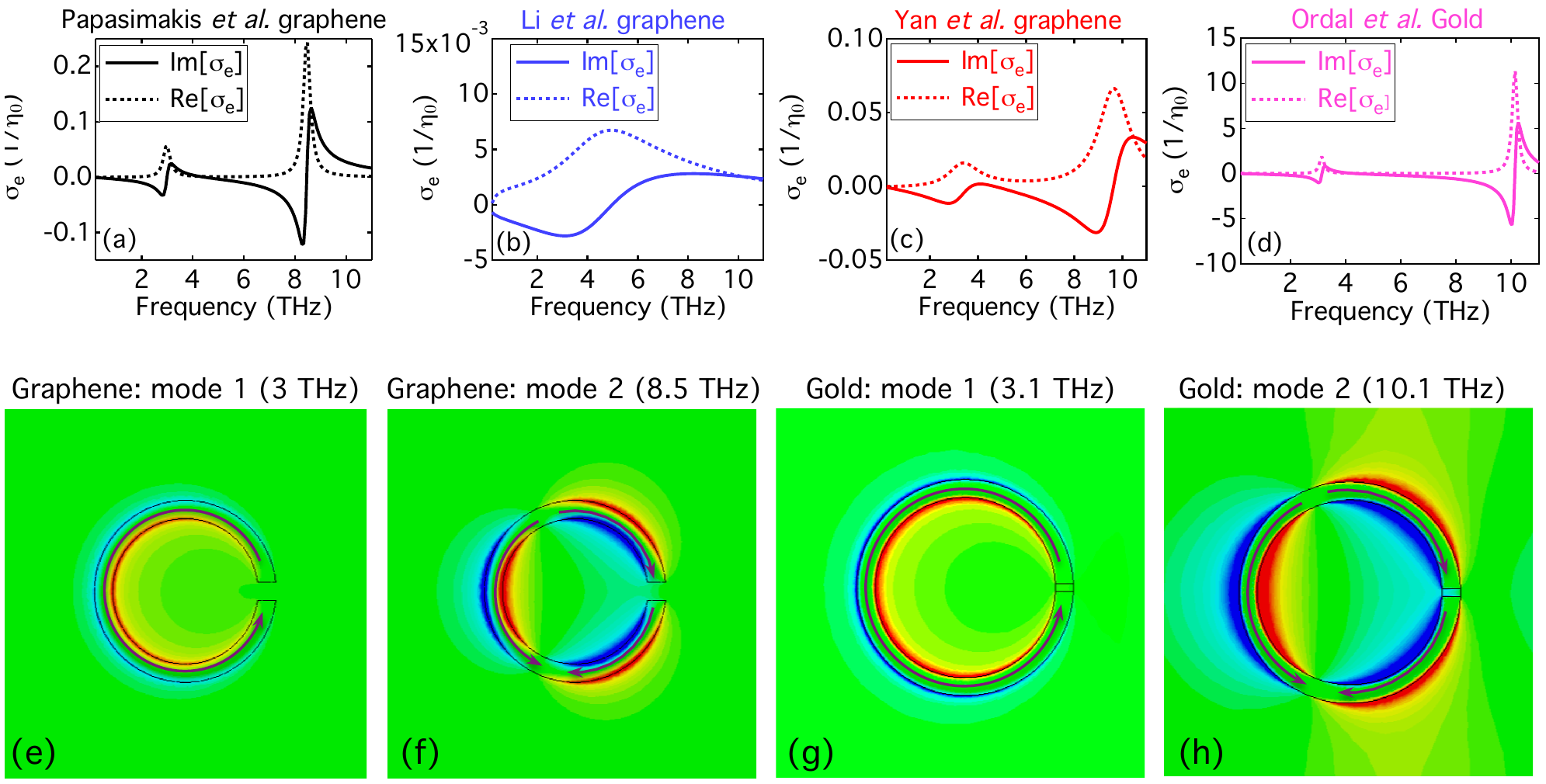}
  \caption{Retrieved frequency dependent electric sheet conductivity (upper row) and $z-$component magnetic field distributions (lower row) for graphene and gold SRRs under normal incidence, respectively. (e) and (f) corresponds two modes for Papasimakis \textit{et al.} graphene case; (g) and (h) are for case of Ordal \textit{et al.} gold SRRs. Arrows in (e)-(h) give the direction of currents.}
\end{figure}

\section{Graphene and gold SRRs under normal incidence}

We first investigate SRRs with normally incident illumination. Figure 1(a) schematically illustrates this SRR configuration. The SRRs are in the $x$-$y$ plane (single layer) and the incident wave propagates under normal incidence with the electric ($E$) and magnetic ($H$) fields polarized along the $y$ and $x$ directions, respectively. The geometric parameters of the graphene SRRs are shown in the caption of Fig. 1. The calculated absorption spectra for the SRRs made out of Yan~\textit{et~al.}, Papasimakis~\textit{et~al.}, and Li~\textit{et~al.} graphene are presented in Fig.~1(b). For the case with Papasimakis~\textit{et~al.} graphene, two fairly sharp absorption peaks are found: the lower-frequency one at \unit{3}{\tera\hertz}, marked as ``$\mathrm{P}_{\mathrm{m}}$,'' comes from the so-called magneto-electric coupling to the magnetic dipole mode, which generates a magnetic dipole along the $z$ direction. This is confirmed by the $z$ component of the magnetic field ($H_z$) shown in Fig.~2(e) where the arrows in the ring demonstrate the circulating current distribution of the magnetic mode of the SRR. The second absorption peak occurring at \unit{8.5}{\tera\hertz}, marked as ``$\mathrm{P}_{\mathrm{e}}$,'' is due to the electric dipole mode with a snap-shot distribution of $H_z$ shown in Fig.~2(f), where the arrows again denote the direction of the surface current. For the case of Yan~\textit{et~al.} graphene, we also find two absorption peaks corresponding to the same modes (``$\mathrm{Y}_{\mathrm{m}}$'' and ``$\mathrm{Y}_{\mathrm{e}}$'') in the frequency range of interest, but both resonances are now weaker due to higher dissipative loss. We also observe that both resonances are blue-shifted compared to those in the Papasimakis~\textit{et~al.} graphene case. This is because of the higher doping and the resulting lower \textit{kinetic inductance} $L_\mathrm{k}$ of the Yan~\textit{et~al.} graphene samples. When we consider the case of Li~\textit{et~al.} graphene in Fig.~1(b), we observe that the resonances are highly damped due to the high dissipative loss in the sample. The absorption peak ``$\mathrm{L}_{\mathrm{e}}$'' is very shallow with a tiny amplitude of the order of magnitude of $10^{-3}$ and we have checked that it is the electric dipole mode; a spectral feature for the magnetic dipole mode is vaguely seen as a very tiny bump, marked as ``$\mathrm{L}_{\mathrm{m}}$.'' The significant redshift of the resonance frequencies in the Li~\textit{et~al.} graphene case is indeed expected from the smaller $\alpha$ value leading to much higher kinetic inductance. One very appealing finding for all graphene cases is that the magnetic dipole mode, occurring at around \unit{3}{\tera\hertz}, is deeply subwavelength, with a $\lambda/a$ ratio as high as 50, and even for the electric dipole mode at higher frequency, $\lambda/a$ still reaches a value of about 19. This finds its origin in the huge \textit{kinetic inductance} of graphene, which dominates over the geometric inductance in the setup under consideration. This is very interesting for the construction of metamaterials, because it allows to work deep in the effective-medium limit and to avoid periodicity artifacts.

For gold SRRs under normal incidence as illustrated in Fig.~1(c), the kinetic inductance ($1/\alpha$) is much smaller compared to graphene, and it is usually the geometric inductance that dominates. Thus, it is expected that both the magnetic and the electric dipole modes would occur at much higher frequencies if the in-plane dimension were left unaltered. Since the resonance frequency can be estimated by $1/[(L_{\mathrm{g}}+L_{\mathrm{k}})C]$, where $L_{\mathrm{g}}$ is the geometric inductance and $C$ the capacitance, there are two different strategies to achieve the same resonance frequencies as in the graphene case discussed above: the first is to increase the capacitance dramatically so as to compensate the significant difference in kinetic inductance between gold and graphene. This would, however, require a tiny ring gap that would most likely be unachievable in an experiment. The second way is to increase the dimension of the SRRs in order to achieve a much larger magnetic inductance. This is experimentally easier to achieve, but it sacrifices the deep subwavelength dimensions of the metamaterial unit. Here we take the second strategy and the results shown in Figs.~1(c)-(d) and 2(g)-(h) are to be compared with the results for the graphene SRRs. It should be noted that essentially such a comparison it is not very fair because the change in the dimensions of the SRRs leads to significant difference in the coupling strength for both cases. However, the comparison still provides us some guidance towards the performance of graphene- and gold-based metamaterials.

\begin{figure}[htb]
  \includegraphics[width=11 cm]{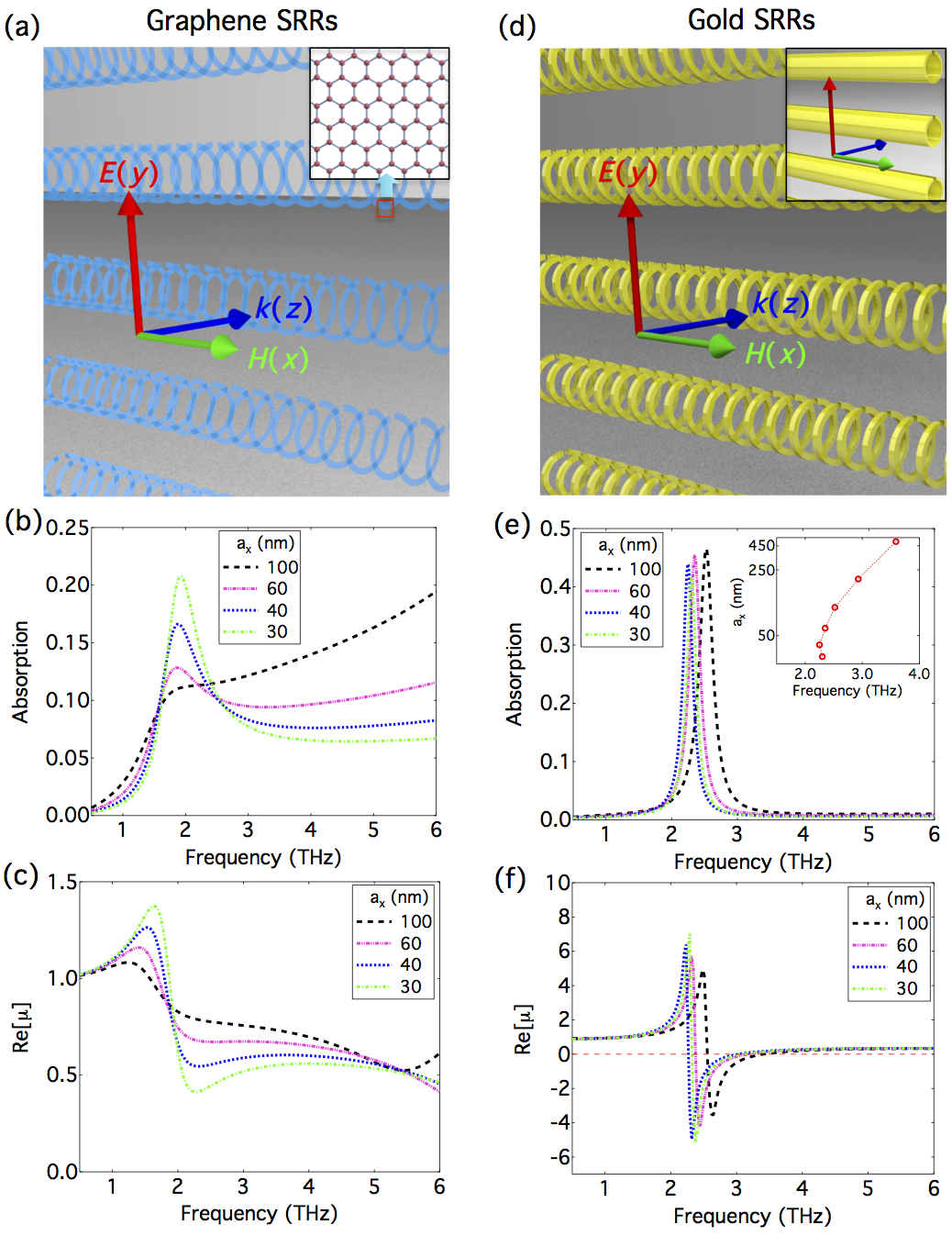}
  \caption{Graphene SRRs [(a)-(c)] and gold SRRs [(d)-(f)] under parallel-incidence illumination. (a) and (d) Schematics of the geometries. Geometric parameters are: outer diameter of the ring $D=\unit{5}{\micro\meter}$, ring width $w=500$ nm, gap size $g=500$ nm, and in-plane lattice constant $a=\unit{6}{\micro\meter}$. (b) Absorption spectra and (c) effective parameter Re[$\mu$] for SRRs made from Yan \textit{et al.} graphene with different ring separations $a_x$: 100, 60, 40, and 30 nm. (e) Absorption spectra and (f) effective parameter Re[$\mu$] for Ordal \textit{et al.} gold SRRs. The inset in (e) illustrates resonance frequency in dependence of ring separation $a_x$ for gold SRRs.}
\end{figure}

The geometry of the gold SRRs, schematically shown in Fig. 1(c), is defined with the parameters shown in the caption of Fig. 1. Benefiting from the increased in-plane dimension, the geometric inductance of the SRRs is able to compensate the difference in kinetic inductance between graphene and gold and the resonance frequencies for the lowest two modes, i.e., the magnetic dipole mode ($\mathrm{G}_{\mathrm{m}}$) and the electric dipole mode ($\mathrm{G}_{\mathrm{e}}$), are 3.1 and \unit{10.1}{\tera\hertz} [see the absorption spectrum in Fig.~1(d)], comparable to those of the graphene SRR metamaterials. Due to the low dissipative loss in gold, we see that the two resonances of the gold SRRs are stronger than those in the graphene cases. This is confirmed by the retrieved effective electric sheet conductivity in Figs.~2(a)-(d). The field distributions of $H_z$ together with arrow plots of the current distribution at the two absorption peaks in Fig.~1(d) are shown in Figs.~2(g) and (h), respectively. They confirm the nature of the two resonant modes. From the above comparison between graphene- and gold-based SRRs under normal incidence, we can conclude that graphene metamaterials are superior in the property of deep sub-wavelengthness due to the huge kinetic inductance of graphene. However, the high dissipative loss in graphene dampens the strength of the resonances even with the high-quality graphene by Yan and co-workers. Nevertheless, the electric dipole resonance of graphene metamaterials is still surprisingly strong and is potentially useful in the applications of terahertz wave manipulation (see the discussions in Part V).

\section{Graphene and gold SRRs under parallel incidence}

Subsequently, we consider graphene-based and gold-based SRRs under parallel incidence, as schematically illustrated in Figs.~3(a) and (d), respectively. The rings are embedded in a dielectric spacer made from a polymer \cite{Farmer2009NanoLett,Yan2012NatNano} with dielectric constant $\epsilon_\mathrm{s}=2.4$ and oriented to avoid magneto-electric coupling, so that we can focus on the magnetic response purely induced by the external magnetic field. For graphene, we use only the experimental data by Yan~\textit{et~al.} in this section, since they are more realistic than theoretical models and bear much lower loss than Li~\textit{et~al.} data. To increase the strength of the magnetic resonance to some extent, we adopt larger SRRs with diameter $D=\unit{5}{\micro\meter}$, lattice constant $a=\unit{6}{\micro\meter}$, ring width $w=\unit{500}{\nano\meter}$ and gap size $g=\unit{500}{\nano\meter}$ for both the graphene and gold cases. The increased area of the graphene SRR results in a stronger induced electromotive force and, hence, in a larger magnetic moment and stronger resonances. Indeed, the non-planar configuration (under parallel incidence), provides a fairer comparison between graphene and gold SRRs, since this geometry allows having both types of SRRs with approximately the same dimensions, so that the coupling strength to the magnetic dipole mode, which is proportional to the area of the SRR, is also approximately the same for both. In Fig.~3, we plot the magnetic response for several separation distances between the rings ($a_x$ varying from \unit{100}{\nano\meter} down to \unit{30}{\nano\meter}), which is the physical limit for the gold case, i.e., the separated gold SRRs become ``split tubes'' when $a_x$ decreases to \unit{30}{\nano\meter} [see Fig. 3(d) and its inset]. Figures~3(b) and (c) show the absorption spectrum and retrieved effective permeability of Yan~\textit{et~al.} graphene SRRs. With decreasing $a_x$, the magnetic resonance of the graphene SRRs is gradually strengthened, but even for $a_x=\unit{30}{\nano\meter}$, it is still not strong enough for the effective permeability to reach negative values. One may notice that the resonance frequency (under parallel illumination) is close to that in the normal-incidence case, even though the rings have much smaller dimensions. This is due to an increased geometric inductance to compensate for the altered capacitance in the case of parallel incidence on the stack of graphene SRRs.

In comparison, the magnetic resonance of gold SRRs is very strong, rendering sharp absorption peaks [Fig. 3(e)] and negative permeability [Fig. 3(f)] for all cases with $a_x$ decreasing from 100 to \unit{30}{\nano\meter}. We find that, due to the densely packed SRRs along the direction of magnetic field ($x$ axis), the geometric inductance is dramatically increased and dominates over the kinetic inductance. Therefore, the geometric inductance and distributed capacitance of the SRRs determine the frequency of the magnetic resonance. This is confirmed by the inset of Fig.~3(e), which shows the relation between the resonance frequency and the ring separation $a_x$ for gold SRRs. The low dissipative loss in gold helps gold SRRs to exhibit superior performance in resonance strength over the graphene SRRs.

\begin{figure}[htb]
  \includegraphics[width=14 cm]{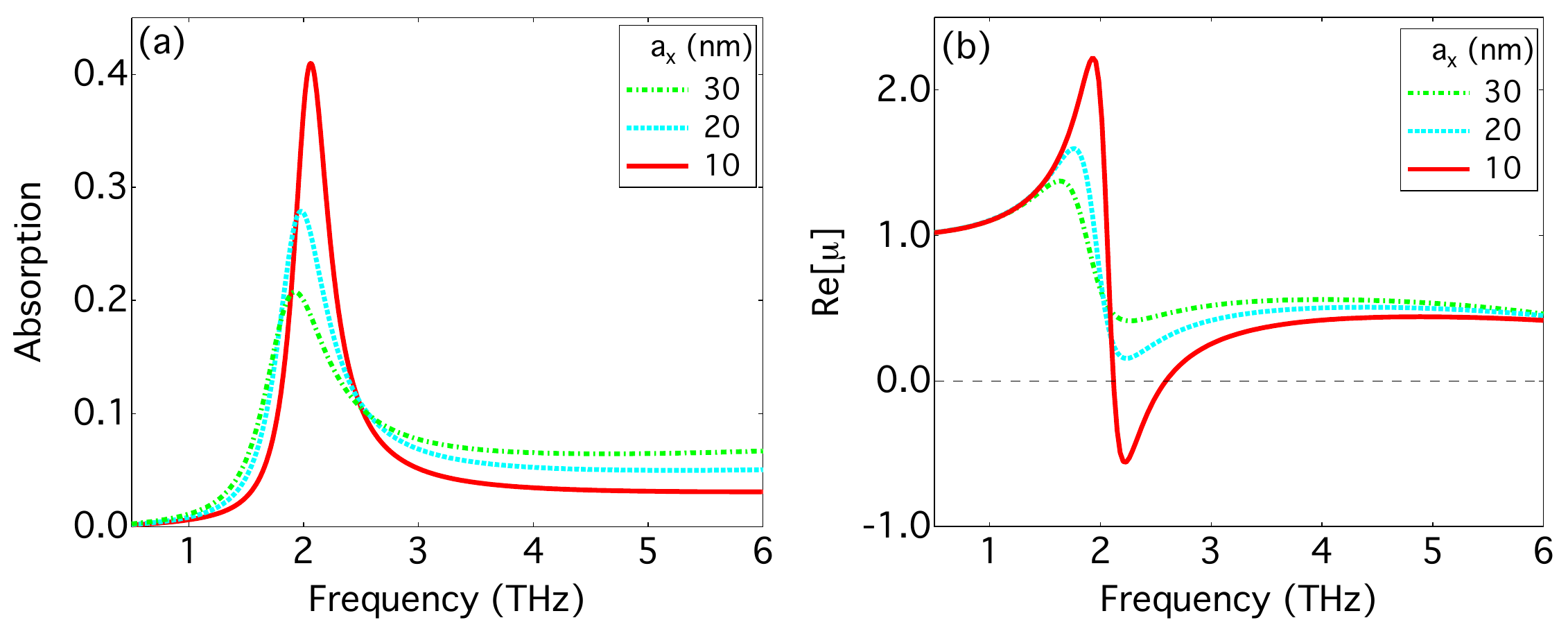}
  \caption{(a) Absorption spectra and (b) effective permeability for Yan \textit{et al.} graphene SRRs under parallel incidence with $a_x$ = 10, 20, 30 nm.}
\end{figure}

However, for the parallel-incidence configuration, we can pack the graphene SRRs even denser, which may lead to further stronger magnetic resonance. Figures~4(a) and (b) show the absorption spectrum and retrieved effective permeability for $a_x$ being \unit{20}{\nano\meter} and \unit{10}{\nano\meter} together with the case for \unit{30}{\nano\meter} from Fig.~3 as a reference. As expected, the strength of the magnetic resonance keeps increasing with decreased $a_x$, and for $a_x = \unit{10}{\nano\meter}$ a negative effective permeability can be achieved. Further decreasing the separation between neighbouring graphene SRRs below \unit{10}{\nano\meter} would make the magnetic resonance even stronger, but it would also severely challenge the fabrication. In Supplementary Note II,\cite{SM} we present the absorption spectra and retrieved effective $\mu$ for SRRs made from several previously listed graphene samples with $a_x = 10$ nm, so from the results, we can see how $\alpha$ and $\gamma$ of graphene determines the performance of SRRs under parallel incidence.

\begin{figure}[htb]
  \includegraphics[width=15.5 cm]{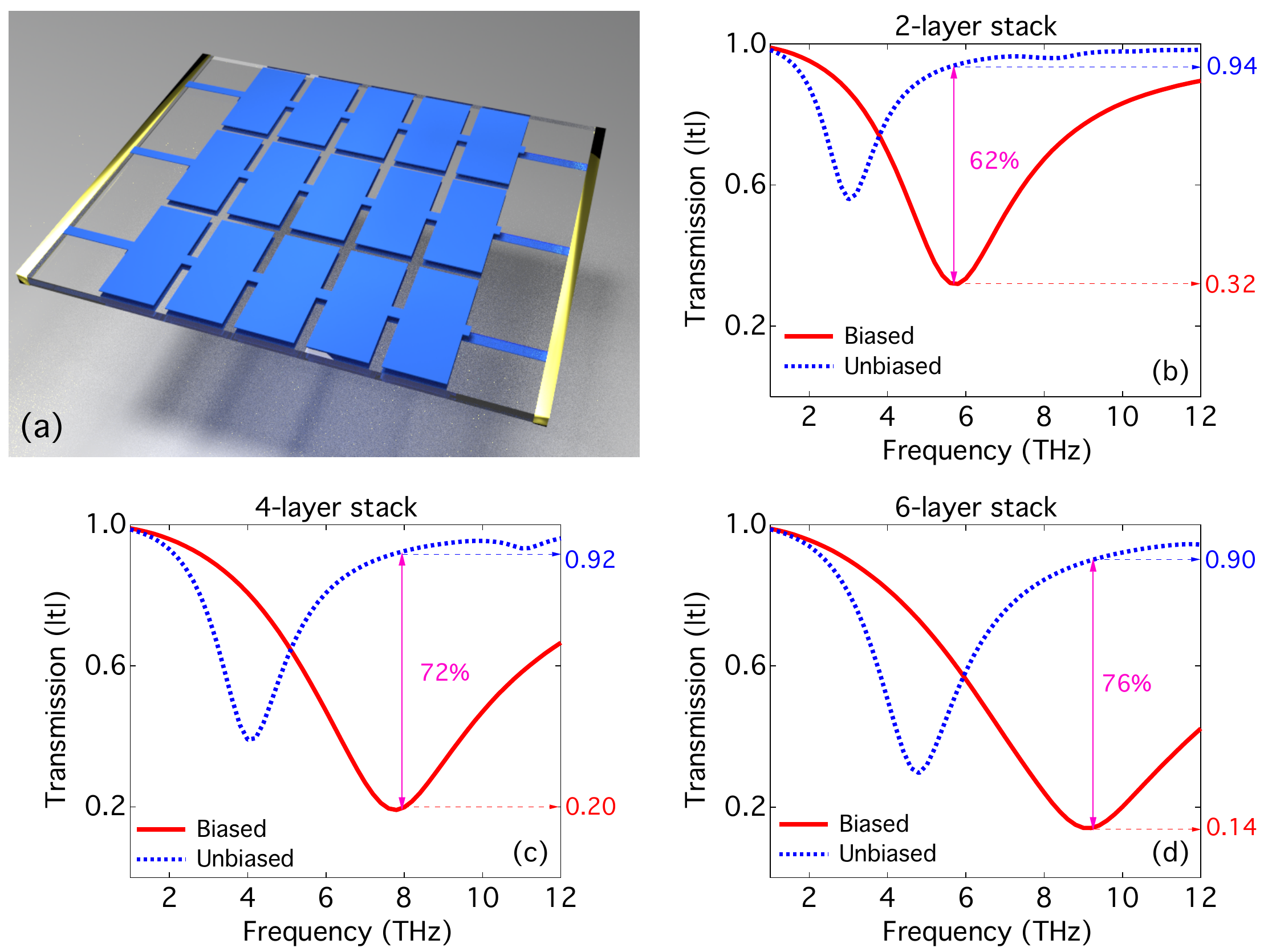}
  \caption{(a) Schematic of prototype of graphene THz switch with two-layer configuration: blue regions represent graphene patterns, transparent spacer is polymer and two side gold bars serve as contact for odd and even graphene layer(s), respectively. (b)-(d) show the performance in transmission for switching on/off through biased/unbiased patterned graphene stack with 2, 4, and 6 layers.}
\end{figure}

\section{Prototype design of graphene terahertz modulator}

So far, we have performed a number of comparisons between graphene and gold SRR metamaterials for both normal- and parallel-incidence geometries (some brief discussions about comparison between graphene and gold cut-wire metamaterials are shown in Supplementary Note III \cite{SM}). We revealed that the huge kinetic inductance of graphene allows to achieve resonant response in the deep sub-wavelength limit under normal incidence when the kinetic inductance dominates.  However, the high dissipative loss of state-of-the-art graphene samples limits the strength of the metamaterial resonances. On the other hand, we should always keep in mind that the most appealing advantage of graphene over noble metals is its tunable electrical properties. In view of our discussions above, we find that graphene metamaterials do show surprisingly strong electric dipole resonances, despite the fairly high dissipative loss in graphene. Therefore, it is advantageous to utilize these tunable electric resonance to create THz modulators. In fact, relating to graphene-aided tunable devices, a lot of efforts have been poured by worldwide researchers. For example, Bludov \textit{et al.} demonstrated a THz switch with a monolayer graphene sheet incorporated in an attenuated total internal reflection structure;\cite{Bludov2010EPL} Gao \textit{et al.} studied the tunable extraordinary optical transmission effect by integrating a graphene sheet to the metallic resonant cavity structure;\cite{Gao2014NanoLett} Sensale-Rodriguez and colleagues explored the modulation effect to a multi-layer configuration of continuous graphene sheets; \cite{Rodriguez2012NatComm} Tamagnone \textit{et al.} theoretically revealed the fundamental limits of a graphene modulator by analyzing the properties of graphene in various frequency bands.\cite{Tamagnone2014NatPhoton} Here, we propose a prototype design of THz switch based on a metamaterial with a multi-layer stack of patterned graphene [shown in Fig. 5(a) for a two-layer configuration], the modulation effect of which will be shown purely due to the resonant property of graphene metamaterial itself. Instead of SRRs, we now pattern graphene films into``cut-wire" constituents, which possess an electric dipole resonance with even better response.

The length of the cut wires is \unit{5.5}{\micro\meter} and their width is \unit{2.5}{\micro\meter}, arranged periodically with lattice constants along the $x$ and $y$ directions being 3 and \unit{6}{\micro\meter}, respectively. The cut wires are connected with each other via \unit{0.5}{\micro\meter}-wide thin strips of graphene. Because of the large inductance of the connecting strips at high frequencies, they will not affect the terahertz response of the cut wires. The patterned graphene layers are stacked together with 20 nm-thick polymer as spacing materials, so that the configurations are compatible with the technology employed in the experimental work of Refs. 23 and 36. In our study, the polymer spacer has dielectric constant $\epsilon_\mathrm{s} = 2.4$ and some loss is taken into account through a loss tangent $\tan \delta = 7 \times 10^{-3}$. \cite{Perret2008MircoEng} Some further investigations into the influence of spacer loss on our THz modulators are presented in Supplementary Note IV.\cite{SM} The thin graphene strips of even and odd layers are connected to electrodes of opposite voltages, providing the alternating sheets in the graphene stack with electron and hole doping, respectively.

This device avoids the need for complicated top or back electrodes, in this way easing the fabrication. More importantly, benefiting from the multi-layer stack configuration, the design increases the carrier density and conductivity in the system dramatically,\cite{Yan2012NatNano} and therefore, the device shows very satisfying performance in tunability. Figures 5(b)-(d) show the results of simulations for configurations of 2-layer, 4-layer, and 6-layer patterned graphene stack. In the simulations, we still apply the realistic experimental data of Yan \textit{et al.} graphene for the biased case (i.e., graphene being heavily \textit{p-}doped). On the other hand, relating to the data of graphene for the unbiased case, a reasonable estimation is made according to the experimental measurements to graphene in THz by Horng \textit{et al.},\cite{Horng2011PRB} in which, it is shown that, for the hole doping regime, $\gamma$ value is more or less constant independent of gate voltage, but at the electron doping side, $\gamma$ increases approximately by 1/2 at highest doping level comparing to that at around charge-neutral point (lightly doping). Therefore, we take the data set of $(\alpha, \gamma)= (\unit{1.9\times10^{10}}{\reciprocal\ohm\reciprocal\second}, \unit{9.8\times10^{12}}{\reciprocal\second})$ for the unbiased graphene in our study, and for the highest electron doping level, graphene is modeled with $(\alpha, \gamma)= (\unit{7.6\times10^{10}}{\reciprocal\ohm\reciprocal\second}, \unit{14.7\times10^{12}}{\reciprocal\second})$. The simulations show that, the tuning efficiency of a 2-layer stack reaches about $62\%$ and for a device of 6-layer graphene stack with only \unit{100}{\nano\meter} thick, it can modulate over $75\%$. 

\section{Conclusions}

In conclusion, we have compared the performance of graphene and gold when used in the design of metamaterials in the terahertz domain. The huge kinetic inductance of graphene results in promising deep subwavelength metamaterial resonances, but the resonances are relatively weaker due to the higher dissipative loss compared to gold. Densely packed graphene SRRs are found to exhibit quite strong magnetic resonances possibly possessing negative permeability, but their performance is not as good as gold SRRs. However, graphene$-$with its easily tunable electrical properties$-$definitely provides significant advantages for tunable metamaterials over gold, especially in achieving miniaturized switchable devices. We have successfully proposed a terahertz modulator based on a multi-layer patterned graphene stack by controlling the surprisingly strong electric resonances in graphene metamaterials. The device, which shows very good performance, is compatible with state-of-the-art experimental technology. Our results provide important guidance for the development of graphene metamaterials and applications to various miniaturized devices in the terahertz domain.  

\section{Acknowledgement}

Work at Ames Laboratory was partially supported by the U.S.\ Department of Energy, Office of Basic Energy Science, Division of Materials Sciences and Engineering (Ames Laboratory is operated for the U.S.\ Department of Energy by Iowa State University under Contract No. DE-AC02-07CH11358) (simulations) and by European Research Council under the ERC advanced Grant No. 320081 (PHOTOMETA) (theory).

\bibliographystyle{apsrev}

\end{document}